\documentclass[twocolumn,showpacs]{revtex4}

\usepackage{graphicx}%
\usepackage{dcolumn}
\usepackage{amsmath}

\makeatletter
\def\btt#1{\texttt{\@backslashchar#1}}%
\DeclareRobustCommand\bblash{\btt{\@backslashchar}}%
\makeatother

\begin{document}

\preprint{HEP/123-qed}

\title[Short Title]{Average Entropy of a Subsystem from its Average Tsallis Entropy}
\author{ L. C. Malacarne$^1$, R. S. Mendes$^1$ and E. K. Lenzi$^2$}                             %
\affiliation{ $^1$Departamento de F\'\i sica, Universidade
Estadual de Maring\'a, Avenida Colombo 5790, 87020-900,
Maring\'a-PR, Brazil\\ $^2$ Centro Brasileiro de Pesquisas F\'\i
sicas, R. Dr. Xavier Sigaud 150,  22290-180 Rio de Janeiro-RJ,
Brazil }

\date{\today}

\begin{abstract}
In the nonextensive Tsallis scenario, Page's conjecture for the
average entropy of a subsystem[Phys. Rev. Lett. {\bf 71},
1291(1993)] as well as its demonstration are generalized, {\it
i.e.}, when a  pure quantum system, whose Hilbert space dimension
is $mn$, is considered, the average Tsallis entropy of an
$m$-dimensional subsystem is obtained. This demonstration is
expected to be useful to study  systems  where the usual entropy
does not give  satisfactory results.
\end{abstract}

\maketitle

\section{Introdution}

Entropy is one of the most ubiquitous quantities in physics. For
example, the entropy is fundamental in the study of quantum and
classical information theories, applied in recent developments in
telecommunications, computer science and engineering (for a
review, see \cite{Vedral,Vedral2}). In particular, a great effort
has been made to understand quantum entanglement of inseparable
quantum system\cite{Peres,Horodecki}. A traditional example of a
pure entangled state is the Einstein-Podolsky-Rosen singlet
state\cite{Einstein}. Another interesting aspect is to obtain
information about the entropy of a subsystem by studying its
average\cite{Page,Lloyd}  over pure states of the big system, in
unitary Haar measure. For instance, a complete  pure  system can
be identified with a black hole and the radiation field related to
it,  in which case the subsystem is the black hole or
alternatively the radiation field\cite{Page2}.

The  standard entropy and its corresponding  thermostatistics
present serious difficulties  when employed to study systems with
long range interaction, in particular, when we deal with
gravitational
interactions\cite{Salzberg,Tisza,Landsberg,Binney,Robertson}. A
possible way to overcome this kind of difficulty  is considering a
new entropy. As stressed by Lavenda and co-workers, a newly
proposed entropy should  have concavity property\cite{Lavenda}.
Such an entropy was considered by Tsallis\cite{Tsallis}.

The  Tsallis entropy,
\begin{equation}\label{eq9}
  S^{(q)} (p_i) =\frac{1-\sum_i p_i^q}{q-1},
\end{equation}
recovers the usual entropy $S(p_i)=S^{(1)}(p_i)=-\sum_i p_i \ln
p_i$ in the limit $q\rightarrow 1$ and has a definite concavity
for all $q$ values ($S^{(q)}$ is concave for $q>0$ and convex for
$q<0$).  Furthermore, if we consider two independent subsystems,
$A$ and $B$, we have  the probabilities $p_{ij}^{A B}= p_i^A
p_j^B$, and
\begin{equation}\label{aditive}
S^{(q)}_{A B} =  S^{(q)}_{A } + S^{(q)}_{ B} +(1-q) S^{(q)}_{A }
S^{(q)}_{ B},
\end{equation}
in contrast with the extensive property of the usual entropy,
$S_{A B} =  S_{A } + S_{ B}$. Thus, the parameter $q$ gives a
measure of the nonextensivity induced by the Tsallis entropy. In
this context, it is common to employ the jargon ``nonextensive" to
refer to the scenario when the Tsallis entropy is present.

Many investigations based on the Tsallis entropy have been
developed. A representative set of such developments  relates to
self-gravitating systems\cite{prl1721}, cosmic background
radiation\cite{prl22},  peculiar velocities in
galaxies\cite{prl23}, L\'evy-type anomalous
superdiffusion\cite{Levy}, H theorem\cite{Plastino},
turbulence\cite{turbo}, nonlinear anomalous diffusion\cite{Ervin},
perturbation and variational methods\cite{mala2}, linear response
theory\cite{29}, Green's functions\cite{30}, and quantum
entanglement\cite{entangle} (for a recent review  see Ref.
\cite{review}).

Since the  Tsallis entropy has played a central role in a
nonextensive scenario such as those cited previously,  it is
natural to investigate  this generalized entropy further. A
different reason for investigating the Tsallis entropy, $S^{(q)}$,
is to technically sneak up on ordinary entropy $S$, yet avoiding
its annoying logarithm by exploiting the $q\rightarrow 1$ limit.
In any case, the aim of this work is to obtain the Tsallis entropy
of a subsystem  averaged over all pure states of the total system
using unitary Haar measure to define our averaging. This result
generalizes Page's conjecture\cite{Page} ( a formula for that
average of the usual entropy of a subsystem) and its subsequent
demonstration\cite{Ruiz,Kanno}. We note that Page's conjecture for
the average entropy of a subsystem has been applied to investigate
black hole radiation\cite{Page2};  perhaps our generalization can
be useful to study parallel reductions to a subsystem in attempts
to fit data with a Tsallis $q$ distinct from 1.

To present our generalization, it is useful to first review Page's
work. This is performed  in Sec. II. Sec. III is addressed to
calculate the average Tsallis entropy of a subsystem. A summary is
given in the last section.

\section{Average Entropy of a Subsystem}

One way to get  entropy out of a system in a pure quantum state is
by  a coarse graining of dividing the system into two subsystems
and ignoring their correlations. Take the system $AB$ with Hilbert
space dimension $mn$ and normalized density matrix $\rho_{AB}$ and
divide it into two subsystems $A$ and $B$, of dimensions $m$ and
$n$ respectively. The entropy of system $A$ is $S_A =- \mbox{tr}
\rho_A \ln \rho_A$, where the density matrix of the system $A$ is
obtained by taking a partial trace over a total system,  $\rho_A=
\mbox{tr}_B \rho_{AB}$. In the same way, $S_B =- \mbox{tr} \rho_B
\ln \rho_B$, with $\rho_B= \mbox{tr}_A \rho_{AB}$. If the system
$AB$ is in a pure state, then $S_{AB}=0$ and $S_A=S_B$ as a
consequence of the fact that $\rho_A$ and $\rho_B$ have the same
set of nonzero eigenvalues\cite{Araki}. Unless the two systems are
uncorrelated in the quantum sense ($\rho_{AB}= \rho_A \bigotimes
\rho_B$,  in which case $S_A=S_B=0$ ), a full quantum analysis is
necessary in order to obtain  $S_A$ and $S_B$, which can be
cumbersome. Yet it is sometimes easy to calculate the unitary Haar
average entropy of the subsystem $A$ over all pure states of the
total system, $S_{m,n} = \langle S_A \rangle$, and  consequently
also the average information of the subsystem, i.e., the deficit
of average entropy from the maximum, $ I_{m,n}= S_{max}^m -
\langle S_A \rangle$, with $S_{max}^m=S(p_i=1/m)$.

For $m\leq n$, Page showed that
\begin{equation}\label{eq2}
S_{m,n}= \int S(p_i) P(p_1,.....,p_m) dp_1, .....dp_m ,
\end{equation}
where $S(p_i)=-\sum_{i=1}^{m} p_i \ln p_i$, and $P(p_1,.....,p_m)$
is the probability distribution of the eigenvalues of $\rho_A$ for
the random pure states $\rho_{AB}$ of the entire
system\cite{Page,Lloyd},
\begin{widetext}
\begin{equation}\label{eq3}
  P(p_1,...,p_m) dp_1...dp_m = N \delta \left(1- \sum_{l=1}^{m}
  p_l \right) \prod_{1\leq i<j\leq m} (p_i-p_j)^2 \prod_{k=1}^{m}
  p_k^{n-m} dp_k.
\end{equation}
In Eq. (\ref{eq2}), as well as in the following integrals, the
integration limits is $0$ and $\infty$. In the above equation,
$N=1/\int P(p_1,.....,p_m) dp_1, .....dp_m$ is the normalization
constant.

By using the identity $1=(\int r^{n m} e^{-r} dr)/(m n \int r^{n
m-1} e^{-r} dr)$ and  the Polygamma function $\Psi(m n+1)=(\int\ln
r~ r^{n m} e^{-r} dr)/( m n \int r^{n m-1}e^{-r} dr )$, we can
write Eq. (\ref{eq2}) as
\begin{eqnarray}\label{eq2b}
S_{m,n}&=&-\frac{ \int e^{-r} ~r^{mn} \sum_i p_i\ln p_i
~P(p_1,.....,p_m)~ dp_1, .....dp_m dr}{ mn \int  e^{-r}
~r^{mn-1}  ~P(p_1,.....,p_m)~ dp_1, .....dp_m dr}\nonumber \\
&-&\frac{ \int  \ln r ~e^{-r}~ r^{mn} ~P(p_1,.....,p_m) ~dp_1,
.....dp_m dr}{ mn \int_0^\infty  e^{-r} ~r^{mn-1}
~P(p_1,.....,p_m) ~dp_1, .....dp_m dr}+ \Psi (mn+1).
\end{eqnarray}
\end{widetext}
Taking into account that $\sum_i p_i =1$, we can introduce the new
variables $x_i=r p_i$; then, by using the delta function to
evaluate the integral in $r$, we obtain
\begin{equation}\label{eq4}
S_{m,n}= \Psi(mn+1)- \frac{\int S(x_i) Q(x_1,...,x_m) dx_1,
...dx_m }{m n\int  Q(x_1,...,x_m) dx_1, ...dx_m },
\end{equation}
with
\begin{equation}\label{eq5}
  Q(x_1,...x_m) dx_1...dx_m = \prod_{1\leq i<j\leq m} (x_i-x_j)^2 \prod_{k=1}^{m}
 e^{-x_k} x_k^{n-m} dx_k.
\end{equation}

Page conjectured\cite{Page}, and other authors
proved\cite{Ruiz,Kanno}, that the exact result  is
\begin{equation}\label{eq6}
S_{m,n} = \sum_{k=n+1}^{mn} \frac{1}{k} - \frac{m-1}{2n}.
\end{equation}
Page had meanwhile applied this to calculate the information in
black hole radiation\cite{Page2}. It was considered a pure
composite total state with a fixed dimension $mn$,  composed by
the black hole and the radiation. The radiation subsystem has
dimension $m$ and the black hole one has dimension $n$. The
average information in the smaller subsystem (for example if you
have $1\ll m \leq n$) is $I_r = S_{max}^m - \langle S_r \rangle
\approx m/2n$.  If furthermore $m\ll n$, the smaller subsystem is
very nearly maximally mixed, and has very little information in
it. The information increases for higher dimension of the smaller
subsystem.

\section{Average Tsallis Entropy}

In this work,  the above result is generalized to ``the
nonextensive case'' as defined by  replacing the usual entropy
[$S(p_i)$] in Eq. (\ref{eq2}) by the Tsallis entropy
[$S^{(q)}(p_i)$]. After similarly introducing the variables $x_i=r
p_i$ in this generalization of Eq. (\ref{eq2}), we obtain
\begin{equation}\label{eq10}
  S_{m,n}^{(q)} = \frac{1}{q-1} -  \frac{1}{q-1}
  \frac{\Gamma(mn)}{\Gamma(mn+q)} J_{m,n}^{(q)} ,
\end{equation}
where
\begin{equation}\label{eq11}
 J_{m,n}^{(q)} = \frac{\int \sum_{i=1}^m x_i^q Q(x_1,...,x_m)
 dx_1...dx_m}{\int  Q(x_1,...,x_m)
 dx_1...dx_m}.
\end{equation}
This expression can be written as a one-dimensional integral in
terms of the one-point correlation function of a Laguerre ensemble
of complex Hermitian random matrices\cite{Mehta}. By considering
the symmetry of $x_i$ and the van der Monde determinant
$\triangle_m (x)= \prod_{1\leq i<j\leq m} (x_i -x_j)$, Eq.
(\ref{eq11}) reduces to
\begin{equation}\label{eq12}
 J_{m,n}^{(q)} =  \int dx_1\;\; x_1^q \;\;\chi (x_1),
\end{equation}
where
\begin{equation}\label{eq13}
  \chi (x_1) =\frac{m \int |\triangle_m (x)|^2 \;\prod_{k=1}^m
  \mu(x_k)\;\;
 dx_2...dx_m}{ \int |\triangle_m (x)|^2 \;\prod_{k=1}^m
 \mu(x_k)\;\;
 dx_1...dx_m},
\end{equation}
with a weight function $\mu(x)=x^{n-m} e^{-x}$. This integration
gives
\begin{widetext}
\begin{equation}\label{eq14}
 \chi (x_1) =\frac{m!}{(n-1)!} x_1^{n-m} e^{-x} \left\{ \left[ L_{m-1}^{n-m+1}
(x_1) \right]^2 -  L_{m-2}^{n-m+1} (x_1) L_{m}^{n-m+1}
(x_1)\right\},
\end{equation}
where $L_r^\alpha (x)$ are the associated Laguerre
polynomials\cite{Mehta} (see also Ref. \cite{Ruiz}).

The remaining integration  in $J_{m,n}^{(q)}$, Eq. (\ref{eq12}),
can be evaluated by taking the following result\cite{integral}:

\begin{equation}\label{eq15}
  \int_0^\infty  x^\theta e^{-x} L_r^\alpha (x) L_s^\beta (x) dx =
 \sum_{k=0}^{min(r,s)} (-1)^{r+s} \left(
\begin{array}{c}
  \theta-\alpha \\
  r-k
\end{array}\right)\left( \begin{array}{c}
  \theta-\beta \\
  s-k
\end{array}\right) \frac{\Gamma(\theta+k+1)}{k!},
\end{equation}
where $\theta >-1$,  $\alpha$ and $\beta$ are real parameters; and
the brackets are binomial coefficients whose factorials of
non-integers or integers $\leq 0$ are interpreted through the
usual $z!=\Gamma(z+1)$.

We finally get  to our goal, a computationally explicit
generalization of Page's conjecture as well as its demonstration,
{\it i.e.},
\begin{eqnarray}\label{Smn}
  S_{m,n}^{(q)} &=&\frac{1}{q-1}- \frac{1}{q-1}
  \frac{\Gamma(m+1)\Gamma(mn)}{\Gamma(n)\Gamma(mn+q)}
 \left[ \sum_{k=0}^{m-1}\left( \begin{array}{c}
 q-1 \\
  m-1-k
\end{array}\right)^2 \frac{\Gamma(n-m+q+1+k)}{k!} \right.
\nonumber \\ &- & \left. \sum_{k=0}^{m-2}\left( \begin{array}{c}
 q-1 \\
  m-2-k
\end{array}\right) \left( \begin{array}{c}
 q-1 \\
  m-k
\end{array}\right) \frac{\Gamma(n-m+q+1+k)}{k!} \right]\; ,
\end{eqnarray}
\end{widetext}
for $m\leq n$.

In the following,  we discuss $S_{mn}^{(q)}$, mainly its
dependence on $q$. Note that  Page's result, Eq. (\ref{eq6}), is
recovered from $S_{mn}^{(q)}$ by taking the appropriate limit
($q\rightarrow 1$), {\it i.e.},  in this limit, Eq. (\ref{Smn})
reduces to
\begin{widetext}
\begin{eqnarray}\label{eq16}
  S_{m,n}^{(q\rightarrow 1)} &=& \Psi(mn+1) -\frac{\Gamma(m+1)\Gamma(mn)}{\Gamma(n)\Gamma(mn+1)}
  \left[ \sum_{k=0}^{m-1} \frac{\Gamma(n-m+2+k)}{[\Gamma(m-k)
  \Gamma(k-m+2)]^2 k!}\right.
  \nonumber \\
  &\times &\left. \left(\frac{}{} 2 \Psi(1) - 2 \Psi (k-m+2)
  + \Psi(n-m+2+k)\right)\frac{}{}\right]
  \nonumber\\
  &+&\frac{\Gamma(m+1)\Gamma(mn)}{\Gamma(n)\Gamma(mn+1)}
  \left[ \sum_{k=0}^{m-2} \frac{\Gamma(n-m+2+k)}{\Gamma(m-k+1)
  \Gamma(k-m+1)\Gamma(m-k-1)\Gamma(k-m+3) k!}\right.
  \nonumber \\
  &\times & \left.\left(\frac{}{} 2 \Psi(1) -  \Psi (k-m+1)-\Psi(3-m+k)
  + \Psi(n-m+2+k)\right)\right]\; .
  \end{eqnarray}
\end{widetext}
In the above equation, the only non-vanishing term in the
summation is that one corresponding to $k$ maximum, so that we
obtain $S_{m,n}^{(q\rightarrow 1)} = \Psi(nm +1) -
\Psi(n+1)-(m-1)/2n.$ By using the relation $\Psi(n+1)=
\sum_{k=0}^{n} 1/k - \gamma$, where $\gamma$ is the Euler's
constant, we get  Page's results, Eq. (\ref{eq6}).

Furthermore, as in the case $q=1$, $S_{mn}^{(q)}$ also assumes  a
simple form when $q$ is a positive integer. This is a consequence
of poles of the $\Gamma(x)$ function for negative integers $x$.
Thus, in the cases of $q=2,3,4,....$, Eq. (\ref{Smn}) reduces to
\begin{widetext}
\begin{eqnarray}\label{SmnInteiro}
S_{m,n}^{(q)} &=&\frac{1}{q-1}- \frac{1}{q-1}
\frac{\Gamma(m+1)\Gamma(mn)}{\Gamma(n)\Gamma(mn+q)} \left[
\sum_{k=1}^{q}\left( \frac{\Gamma (q)}{\Gamma (k) \Gamma (q+1-k)
}\right)^2 \frac{\Gamma(n+q+1-k)}{(m-k)!} \right. \nonumber \\ &-
& \left. \sum_{k=1}^{q-2}\left( \frac{\Gamma (q)}{\Gamma (k)
\Gamma (q+1-k) }\right) \left(\frac{\Gamma (q)}{\Gamma (2+k)
\Gamma (q-1-k) } \right) \frac{\Gamma(n+q-k)}{(m-1-k)!} \right]\;
.
\end{eqnarray}
\end{widetext}
Note that  the second sum only gives contribution for
$q=3,4,5,...$. In particular, for $q=2$, the Tsallis entropy leads
to the quadratic entropy. This entropy was firstly used in
theoretical physics by Fermi (see p. 31, Eq. 2.11.3 of Ref.
\cite{Fermi}). In this case, Eq. (\ref{SmnInteiro}) reduces to
\cite{Lubkin1}
\begin{equation}\label{eq18}
S_{m,n}^{(q=2)} = 1-\frac{n+m}{m n+1}.
\end{equation}
If we observe that the maximum $q$-entropy, obtained when
$p_i=1/m$, is given by $ S_{max}^{(q)m}= (1-m^{1-q})/(q-1)$, the
average information, $I_{m,n}^{(q)}= S_{max}^{(q)m} -<S_A^{(q)}>$,
for $q=2$ is
\begin{equation}\label{eq20}
  I_{m,n}^{(q=2)}=\left(1-\frac{1}{m}\right) -\left(1-\frac{m+n}{m
  n+1}\right)\approx\frac{1}{n}
\end{equation}
for $ mn \gg 1$. Observe that for $mn\gg 1$, $I_{m,n}^{(q=2)}$ is
a power law with only $n$ dependence. Thus, for a system $AB$ with
fixed $mn$ dimension, a log-log plot of $I_{m,n}^{(q=2)}$ versus
$m$ gives a straight line.

\begin{figure}
 \centering
 \DeclareGraphicsRule{ps}{eps}{*}{}
\includegraphics*[width=8cm, height=5cm,trim=1cm 1cm 0cm 1cm]{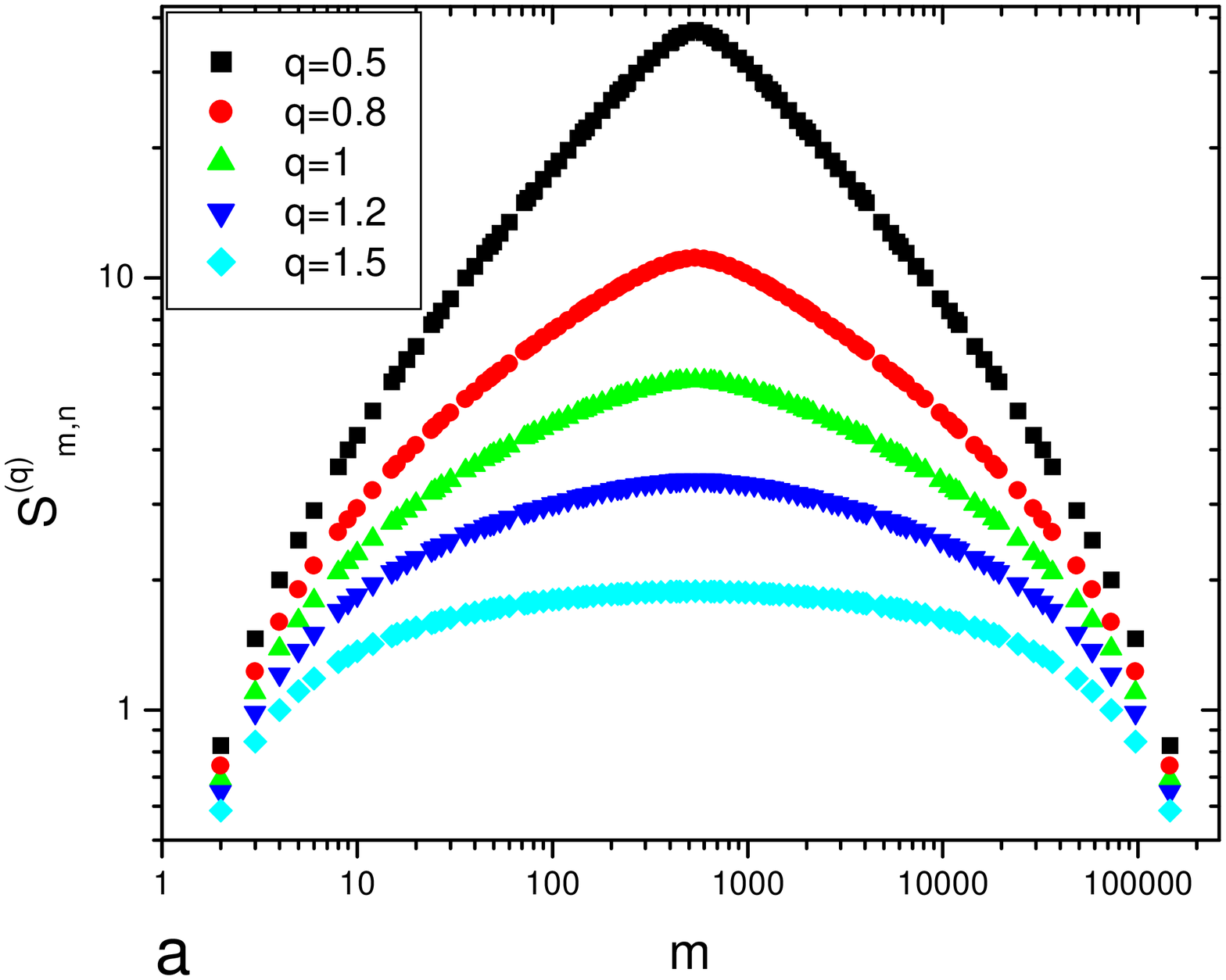}
\includegraphics*[width=8cm, height=5cm,trim=1cm 0cm 0cm 1cm]{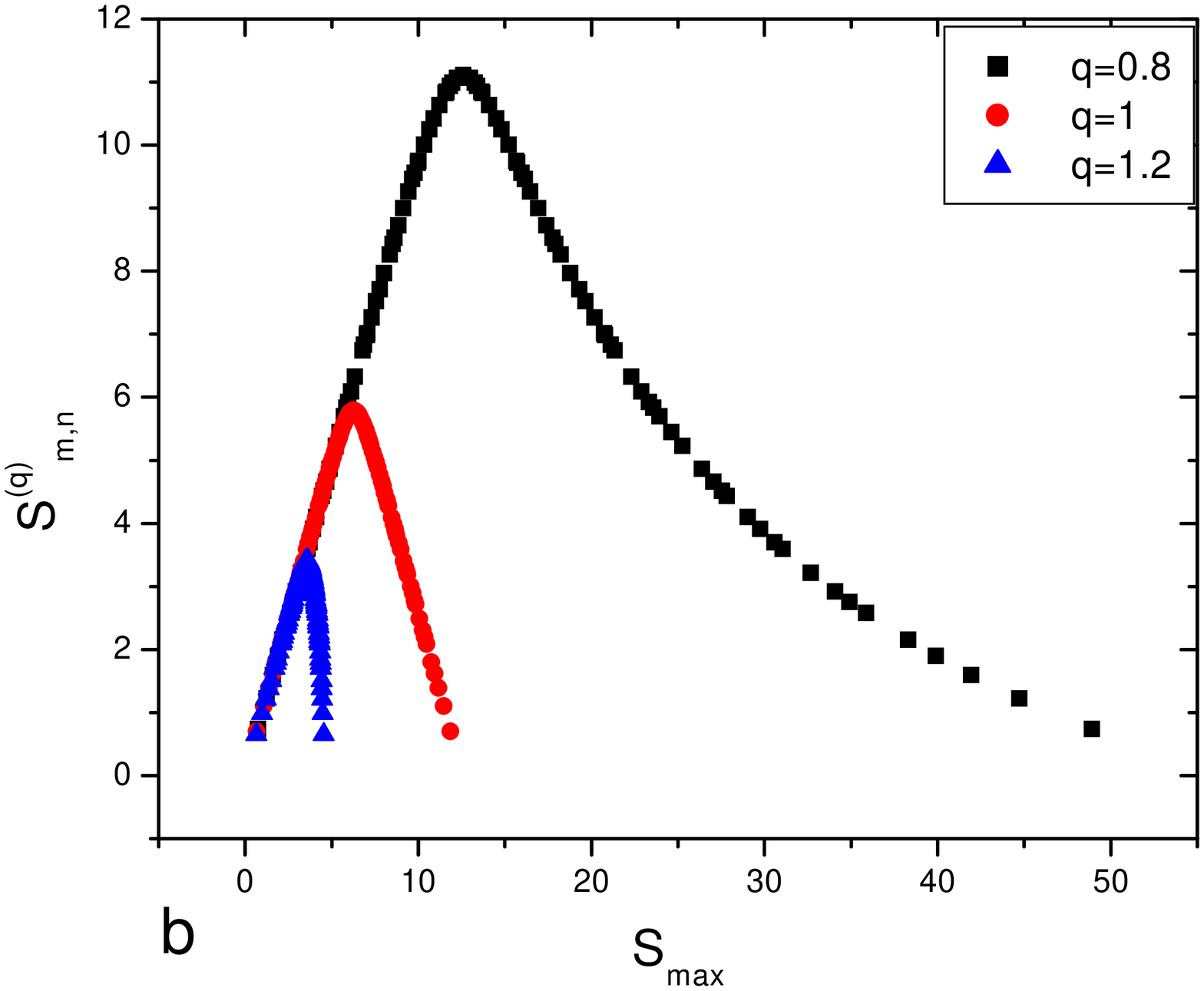}
   \caption{a) Plot of $<S_A^{(q)}>$ versus $m$ to $q=0.5$, $q=0.8$, $q=1$, $q=1.2$ and $q=1.5$ with $mn=291600$.
   b) Plot of $<S_A^{(q)}>$ versus $S_{max}^{(q)}$ to  $q=0.8$, $q=1$ and $q=1.2$ with $mn=291600$.}\label{fig1}
\end{figure}

\begin{figure}
 \centering
 \DeclareGraphicsRule{ps}{eps}{*}{}
\includegraphics*[width=8cm, height=5cm,trim=1cm 1cm 0cm 1cm]{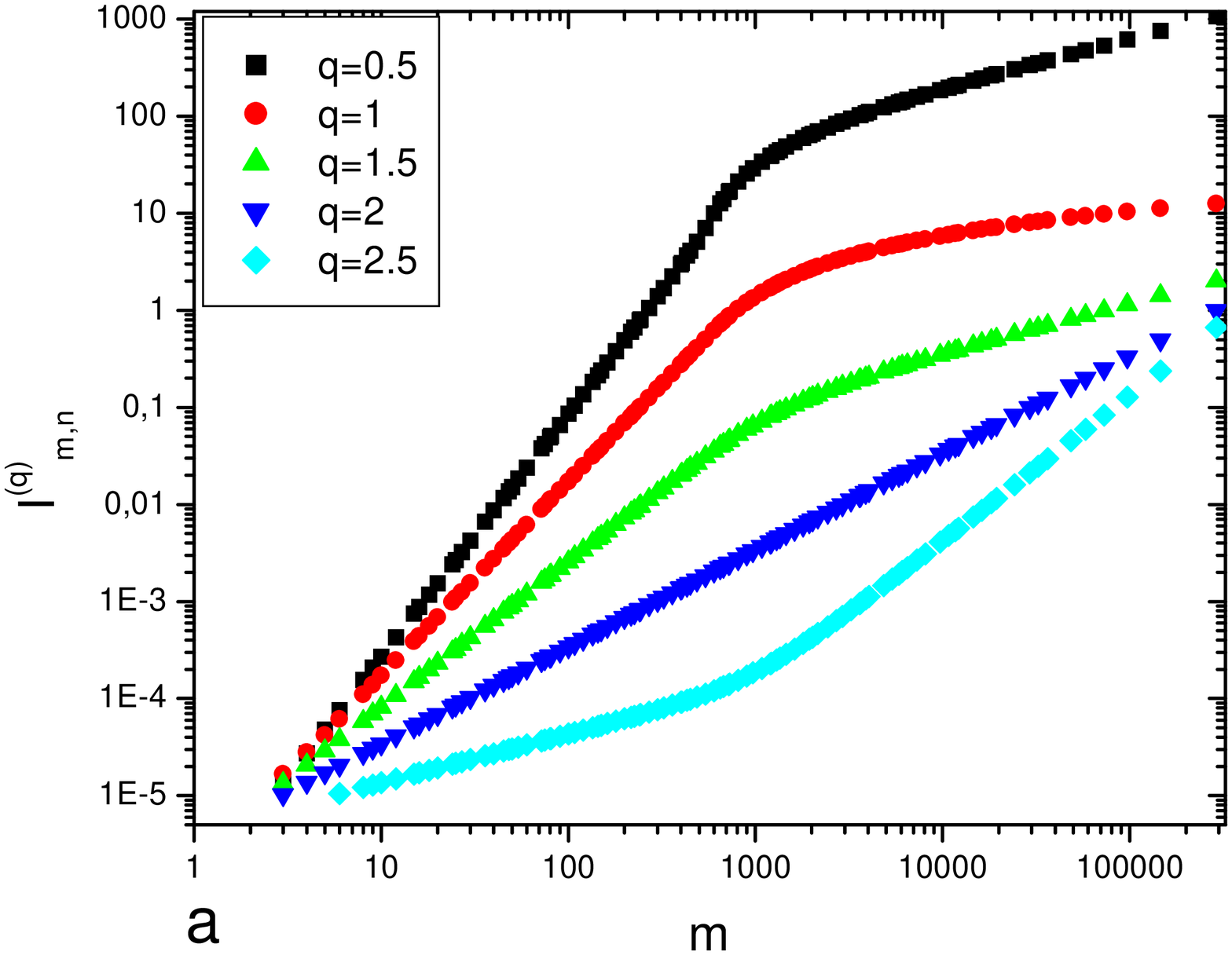}
\includegraphics*[width=8cm, height=5cm,trim=1cm 0cm 0cm 1cm]{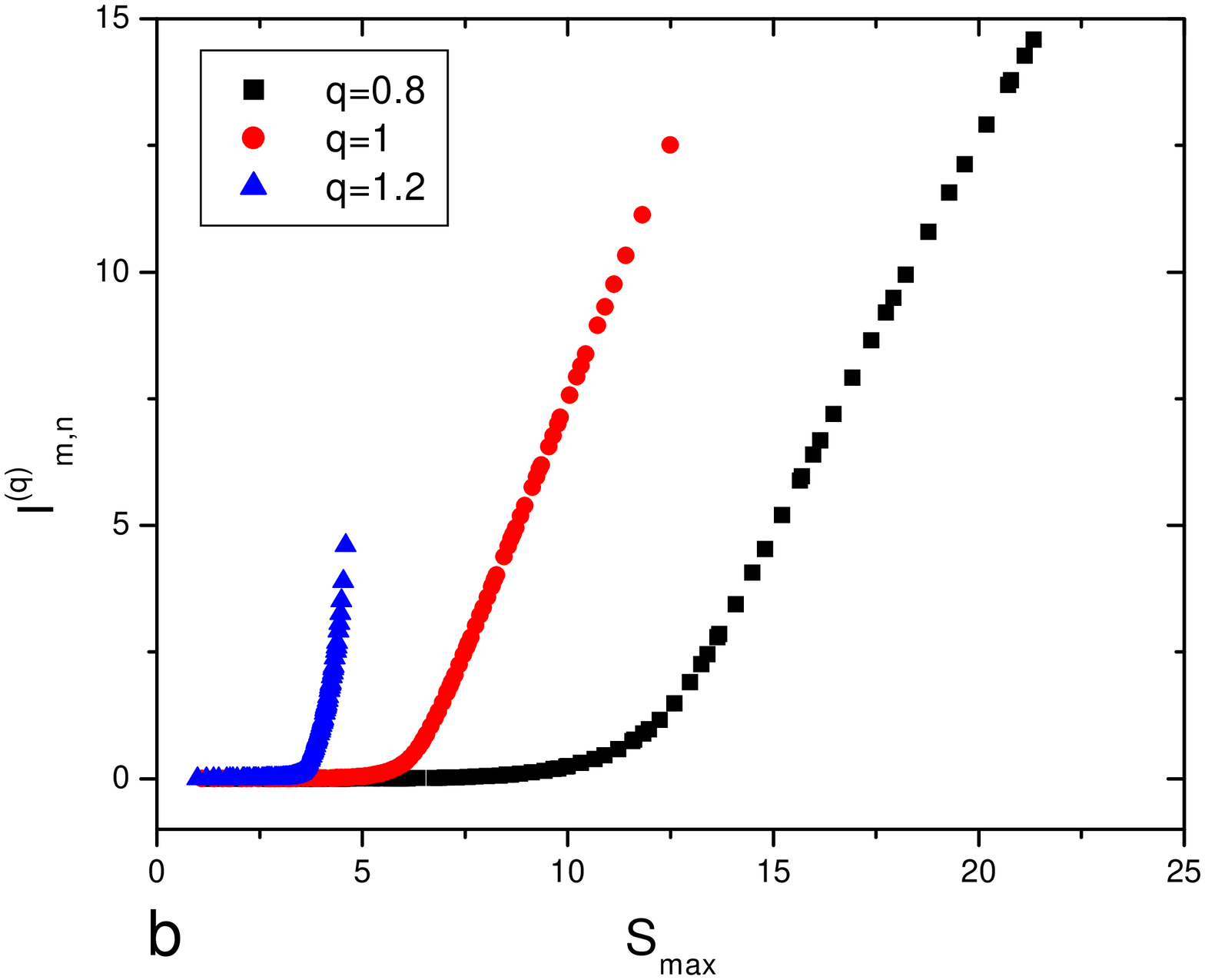}
   \caption{a) Plot of $I_{m,n}^{(q)}$   versus $m$ to  $q=0.5$, $q=1$, $q=1.5$, $q=2$ and
$q=2.5$ with $mn=291600$.   b) Plot of $I_{m,n}^{(q)}$    versus
$S_{max}^{(q)}$ to  $q=0.8$, $q=1$ and $q=1.2$ with
$mn=291600$.}\label{fig2}
\end{figure}

For an arbitrary $q$ value, Eq. (\ref{Smn}) does not reduce to a
simple form,  so we show some graphs instead. For example,
consider a total system with fixed Hilbert space dimension
$mn=291600$ (about the number of states very naively expected for
a black hole near the Planck mass\cite{Page2}). In the case of a
total pure state, we have $<S_A^{(q)}>=<S_B^{(q)}>= S_{m,n}^{(q)}$
if $m\leq n$, and $<S_A^{(q)}>=<S_B^{(q)}>= S_{n,m}^{(q)}$ if $m
\geq n$, where $S_{m,n}^{(q)}$  is given by Eq. (\ref{Smn}) and
$S_{n,m}^{(q)}$ is obtained from it by performing the exchange
$m\leftrightarrow n$. In  Fig. (\ref{fig1}), we plot $<S_A^{(q)}>$
for some representative $q$ values. Fig. (\ref{fig2}) shows the
average information $I_{m,n}^{(q)}$ to different $q$ values.

\section{Summary}

Summing up, we have generalized  Page's conjecture and its
demonstration in order to incorporate the nonextensive regime
induced by the Tsallis entropy. Naturally, this result  must and
does reduce  to the usual one in the limit $q \rightarrow 1$. For
other representative $q$ values  and $mn$ still fixed at 291600,
average entropy and average information  are log-log plotted,
$S^{(q)}$ versus $m$ then $S^{(q)}$ versus $S^{(q)}_{max}$ in Fig.
(\ref{fig1}), and $I^{(q)}$ versus $m$ then $I^{(q)}$ versus
$S^{(q)}_{max}$ in Fig. (\ref{fig2}). The straightness shown by
the triangles in Fig. (\ref{fig2}-a) illustrates the case $q=2$ as
a separation between two different regimes. In general,
calculations based on the nonextensive Tsallis entropy have been
addressed in the study of systems with long range interaction,
spatiotemporal complexity, and fractal structure; thus, we hope
our result may be useful for  such systems.

More  formal applications of the $q\rightarrow 1$ limit to derive
ordinary entropies, may also turn out feasible, for kinds of
averaging other than Haar-unitary, in particular, for time
averaging under Gaussian-distributed Hamiltonians  which do not
discriminate  between $m$ system and $n$ system, and also for
similar distributions which, instead, do discriminate  so as to
model approximate mutual isolation. In both cases, results for
$q=2$ are known \cite{Lubkin2} and the $q \rightarrow 1$ limit
would be welcome. Such further applications would be analogous to
our demonstration in this present paper of Page's conjecture, in
being independent of the issue of whether the Tsallis entropy for
$q\neq1$ is or is not directly applicable to physical situations.

We thank  CNPq (Brazilian Agency) for partial financial support.


\begin{thebibliography}{10}
\bibitem{Vedral} M. B. Plenio, V. Vedral, Contemp. Phys. {\bf 39},
431 (1998).
\bibitem{Vedral2} V. Vedral, {\it The role of Relative Entropy in
Quantum Information Theory}, quant-ph/0102094.
\bibitem{Peres} A. Peres, Phys. Rev. Lett. {\bf 77}, 1413 (1996).
\bibitem{Horodecki} M. Horodecki, P. Horodecki and R. Horodecki, Phys. Rev. Lett.
{\bf 80}, 5239 (1998); Phys. Rev. Lett. {\bf 78}, 574 (1997).
\bibitem{Einstein} A. Einstein, B. Podolsky and N. Rosen, Phys.
Rev. {\bf 47}, 777 (1935).
\bibitem{Page} D. N. Page, Phys. Rev. Lett. {\bf 71}, 1291 (1993).
\bibitem{Lloyd} S. Lloyd and H. Pagels, Annals of Physics {\bf
188}, 186 (1988).
\bibitem{Page2} D. N. Page, Phys. Rev. Lett. {\bf 71}, 3743
(1993).
\bibitem{Salzberg} A. M. Salzberg, J. Math. Phys. (N.Y.) {\bf 6}, 158
(1965).
\bibitem{Tisza} L. Tisza, {\it Generalized Thermodynamics} (MIT
Press, Cambridge, 1966), p. 123.
\bibitem{Landsberg} P. T. Landsberg, J. Stat. Phys. {\bf 35}, 159
(1984).
\bibitem{Binney} J. Binney and S. Tremaine, {\it Galactic
Dynamics} (Princeton University Press, Princeton, 1987), p. 267.
\bibitem{Robertson} H. S. Robertson, {\it Statistical
Thermophysics} (Princeton Prentice-Hall, Englewood Cliffs, NJ,
1993), p. 96.
\bibitem{Lavenda} B. H. Lavenda and J. Dunning-Davies, Found.
Phys. Lett. {\bf 3}, 435 (1990); B. H. Lavenda and J.
Dunning-Davies, Nature (London) {\bf 368}, 284 (1994); B. H.
Lavenda, J. Dunning-Davies and M. Compiani, Nuovo Cimento B {\bf
110}, 433 (1995).
\bibitem{Tsallis}C. Tsallis,
{\it  J. Stat. Phys.} {\bf 52}, 479 (1988);
 E. M. F. Curado and C. Tsallis,
{\it J. Phys. A: Math. Gen.} {\bf 24}, L69 (1991); Corrigenda:
{\bf 24}, 3187 (1991) and {\bf 25}, 1019 (1992);  C. Tsallis, R.
S. Mendes and A. R. Plastino, {\it Physica A} {\bf 261}, 534
(1998).
\bibitem{prl1721} B. M. Boghosian, Phys. Rev. E {\bf 53}, 4754
(1996); A. R. Plastino and A. Plastino, Phys. Lett. A {\bf 174},
384 (1993); J. J. Aly, In {\it Proceedings of N-Body Problems and
Gravitational Dynamics}, Aussois, France, edited by f. Combes and
E. Athanassoula (Plubication de l'Observatoire de Paris, Paris,
1993), p.19; V. H. Hamity and D. E. Barraco, Phys. Rev. Lett. {\bf
76} 4664 (1996).
\bibitem{prl22} C. Tsallis, F. C. S\'a Barreto, and E. D. Loh,
Phys. Rev. E {\bf 52}, 1447 (1995).
\bibitem{prl23} A. Lavagno, G. Kaniadakis, M. Rego-Monteiro, P.
Quarati, and C. Tsallis, Astrophys. Lett. Commun. {\bf 35}, 449
(1998).
\bibitem{Levy} C. Tsallis, S. V. F. Levy, A. M. C. Souza,
and R. Maynard, Phys. Rev. Lett. {\bf 75}, 3589 (1995); {\bf 77},
5442(E) (1996); M. Buiatti, P. Grigolini and A. Montagnini, Phys.
Rev. Lett. {\bf 82}, 3383 (1999).
\bibitem{Plastino} J. A. S. Lima, R. Silva, and A. R. Plastino,
Phys. Rev. Lett. {\bf 86}, 2938 (2001).
\bibitem{turbo} T. Arimitsu and N. Arimitsu, Phys. Rev. E {\bf
61}, 3237 (2000); C. Beck, G. S. Lewis and H. L. Swinney, Phys.
Rev. E {\bf 63}, 035303 (2001).
\bibitem{Ervin} A. R. Plastino and A. Plastino, Physica A {\bf 222}, 347 (1995);
L. Borland, F. Pennini, A. R. Plastino and A. Plastino, Eur. Phys.
J. B  {\bf 12}, 285 (1999); L. C. Malacarne, R. S. Mendes, I. T.
Pedron, and E. K. Lenzi, Phys. Rev. E {\bf 63}, 030101(R) (2001).
\bibitem{mala2}E. K. Lenzi, L. C. Malacarne, R. S. Mendes, Phys.
Rev. Lett. {\bf 80}, 218 (1998); R. S. Mendes, Kwok Sau Fa, E. K.
Lenzi, J. N. Maki, Eur. Phys. J. B {\bf 10},353, (1999); A.
Plastino, C. Tsallis, J. Phys. A {\bf 26}, L893 (1993).
\bibitem{29} A. K. Rajagopal,  Phys. Rev. Lett.  {\bf 76}, 3469 (1996).
\bibitem{30} A. K. Rajagopal, R. S. Mendes,  E. K. Lenzi, Phys.
Rev. Lett.  {\bf 80}, 3907 (1998); E. K. Lenzi, R. S. Mendes, A.
K. Rajagopal, Phys. Rev. E {\bf 59}, 1398 (1999); S. Abe, Eur.
Phys. J. B {\bf 9} 679 (1999).
\bibitem{entangle} S. Abe and A. K.  Rajagopal, Phys. Rev. A {\bf
60}, 3461 (1999); A. Vidiella-Barranco, Phys. Lett. A {\bf 260},
335 (1999); A. Vidiella-Barranco and H. Moya-Cessa, Phys. Lett. A
{\bf 279}, 56 (2001); C. Tsallis, S. Lloyd and  M. Baranger, Phys.
Rev. A {\bf 63}, 042104 (2001).
\bibitem{review} C. Tsallis, in {\it Nonextensive Statistical
Mechanics and Its Applications}, Eds. S. Abe and Y. Okamoto,
(Springer-Verlag, New York, 2001).
\bibitem{Ruiz} J. Sanchez-Ruiz, Phys. Rev. E {\bf 52}, 5653
(1995).
\bibitem{Kanno} S. K. Foong and S. Kanno, Phys. Rev. Lett. {\bf
72}, 1148 (1994); S. Sen, Phys. Rev. Lett. {\bf 77}, 1 (1996).
\bibitem{Araki} H. Araki and E. H. Lieb, Commun. Math. Phys. {\bf
18}, 160 (1970).
\bibitem{Mehta} M. L. Mehta, {\it Random Matrices}, 2nd ed.
(Academic Press, New York, 1991).
\bibitem{integral}
This kind of integrals involving Laguerre Polynomials  appears in
the study of quantum mechanics systems. For instance, see: E.
Schrodinger, Ann. Phys. (Leipzig) {\bf 79}, 361 (1926);  {\bf 79},
489 (1926); {\bf 80}, 427 (1926); {\bf 81}, 109 (1926); M. M.
Nieto and L. M. Simmons, Jr., Phys. Rev. A {\bf 19}, 438 (1979);
M. M. Nieto, Am. J. Phys. {\bf 47}, 1067 (1979); J. S. Dehesa, R.
J. Y\'a\~nez, A. I. Aptekarev, and V. Buyarov, J.  Math. Phys.
{\bf 39}, 3050, (1998); H. A. Mavromatis, R. S. Alassar, App.
Math. Lett. {\bf 14}, 903 (2001); P. M. Morse and H. Feshbach,
{\it Methods of Mathematical Physics}, Vol. 1. p. 785, McGraw-Hill
(1953). The specific form employed here, Eq. (\ref{eq15}), is
given in Eq. (17) of  Ref. \cite{Ruiz}.
\bibitem{Fermi} G. Jumarie, {\it Relative Information} (Springer,
Berlin-1990).



\bibitem{Lubkin1} E. Lubkin, J. Math. Phys. {\bf 19}, 1028 (1978).

\bibitem{Lubkin2} E. Lubkin and T. Lubkin, Int. J. Theor. Phys.
{\bf 32}, 933 (1993).
\end{thebibliography}

\end{document}